# A structural modeling approach to solid solutions based on the similar atomic environment

Fuyang Tian [1a], De-Ye Lin [2,3a], Xingyu Gao [3*], Ya-Fan Zhao [2,3†], and Hai-Feng Song [3,2‡]

1. Institute for Applied Physics, University of Science and Technology Beijing, Beijing 100083, China
2. CAEP Software Center for High Performance Numerical Simulation, Beijing 100088, China
3. Institute of Applied Physics and Computational Mathematics, Beijing 100088, China

**Abstract** Solid solution is an important way to enhance the structural and functional performances of materials. In this work, we develop a structural modeling approach to solid solutions based on the similar atomic environment (SAE). We propose the similarity function associated with any type of atom cluster to describe quantitatively the configurational deviation from the desired solid solution structure that is fully disordered or contains short-range order (SRO). In this manner, the structural modeling for solid solution is transferred to a minimization problem in the configuration space. Moreover, we pay efforts to enhance the practicality and functionality of this approach. The approach and implementation are demonstrated by the cross-validations with the special quasi-random structure (SQS) method. We apply the SAE method to the typical quinary CoCrFeMnNi high-entropy alloy, continuous binary Ta-W alloy and ternary CoCrNi medium-entropy alloy with SRO as prototypes. In combination with *ab initio* calculations, we investigate the structural properties and compare the calculation results with experiments.

## 1. Introduction

The solid solution is a typical disordered system where atomic positions are arranged by the underlying lattice, but chemical elements in the lattice are randomly distributed. It composes of at least two kinds of chemical elements and adopts a typical crystal structure, such as face centered cubic (fcc), body centered cubic (bcc), hexagonal close packed (hcp), L1$_2$, L2$_1$, B2, NaCl, diamond and perovskite structures, etc. Solid solution is one of the fundamental methods to improve the mechanical strength, corrosion and oxidation resistance of structural materials and to modulate energy band of functional materials. For example, the Cr element (>13%) is responsible for the resistance against corrosion of stainless steel in various chemical environments [1]. The partial solid solution on sublattice site in L1$_2$ phase is helpful to regulate Co$_3$(Al,W) as potential high-temperature structural material [2]. More recently, the high-entropy materials are basically solid solutions, including high-entropy alloy (HEA), high-entropy metallic glasses, high-entropy ceramics, high-entropy thermoelectric materials, high-entropy oxides, and so on [3].

[*] gao_xingyu@iapcm.ac.cn
[†] zhao_yafan@iapcm.ac.cn;
[‡] song_haifeng@iapcm.ac.cn;

The electronic structure, phase stability, and intrinsically structural and functional properties of solid solutions could be obtained by using *ab initio* calculations based on density functional theory [8,9]. Recently, the *ab initio*-based high-throughput calculation or material gene engineering has been launched to accelerate the development of novel materials [4–7]. An *ab initio* calculation for materials usually starts with the structural modeling. Whereas, extending *ab initio* calculation to the multicomponent system in the presence of chemical disorder, remains a challenging problem, due to the breakdown of translational symmetry and a large degree of uncertainty on the configuration. Two main categories of methods have been developed to study the phases of chemically disordered solid-solution. One type is the effective medium theory, including the virtual crystal approximation (VCA) [10] and the coherent potential approximation (CPA) [11]. Another type is the finite-size supercell method, including the cluster expansion (CE) method [12], special quasi-random structure (SQS) method [13], and small set of ordered structures (SSOS) method [14]. The VCA adopts the oversimplified average of corresponding one-electron potential of chemical elements. However, neither the electron potential nor the wave function are self-averaging quantities [15]. The CPA method based on the mean-field theory can elegantly treat both chemical and magnetic disorders in fully disordered alloys at arbitrary composition, while the fluctuation of atomic environment is ignored [16]. The SQS method is an approach to modeling the fully disordered solid solution with the best possible small periodic supercell based on the CE theory. The SSOS is expected to use the finite SQS configurations to simulate the disordered structure of multicomponent alloys.

In solid solution, different atom sizes and bonding behaviors of chemical elements derived from solute elements produce unexpected local environments of solute atoms and solvent atoms, which may induce local lattice distortions and short-range order (SRO). These local environments may result in excellent properties. For example, the hardness of an alloy often increases with increasing atomic radius difference of chemical elements [17]. The SRO may play an important role in the low electronic and thermal conductivity of the equimolar CoCrNi solid-solution medium entropy alloy [18,19]. The different bonding behaviors of chemical elements may induce partial disorder in solid solution. For instance, L2$_1$ (NiCo)$_2$TiAl Heusler phase enhances creep resistance in multi-phase alloys [20,21] and the L1$_2$ Co$_3$(Al,W) is a potential high-temperature structural material [2,22]. With increase of Al content, the paramagnetic CoCrFeNiAl$_x$ ($x$=0-2) adopts the fcc structure ($x < 0.60$) and the bcc structure ($x > 1.23$), with an fcc-bcc duplex region between the two pure phases at room temperature [23–25]. For the experimental investigation of solid solution, the local lattice distortion and elastic modulus misfit are often measured via the pair distribution function (PDF) of atoms which represents the local coordination of an atom with its neighboring atoms [26]. However, as far as we know, there are few feasible *ab initio* calculation methods that could take into account the SRO effect properly, in principle, although the CE method can capture the SRO effect in solid solution [27]. Besides, the SRO could not be considered directly in the SQS implementation of the popular package

ATAT (Alloy Theoretic Automated Toolkit), and the modeling efficiency decreases significantly as the supercell size increases [27].

In this work, we propose a finite-size supercell-based modeling approach to the random solid solutions. We start from the assumption on the similar atomic environment (SAE) of the equivalent lattice sites in the sense of statistical average. Actually, it is infeasible to simultaneously optimize the each individual atomic environment in a finite-size supercell. Instead, the overall atomic environment associated with the sublattice is more appropriate for finite-size modeling. The statistical average of overall atomic environments is very similar to that of the fully disordered structure, which is the underlying idea of our approach. The expectation of the overall atomic environment is used to establish the similarity function which measures the deviations from the desired fully disordered or short-range ordered (SRO) structure. In this way, the structural modeling turns into a minimization problem in the configuration space. It will be shown that such a similarity function could be regarded as a generalized SRO parameter in terms of any type of atom cluster. Consequently, the SRO can be naturally considered in the process of the SAE method.

The SAE method should be distinguished from the SQS method in which a cluster function is used to formulate the site occupation correlation function. Alternatively, the similarity function is presented to realize the correlation function in the SAE method. It will be proved that the optimal solution of the SAE method suffices as the optimal solution of the SQS method.

In the section of Methodology, we introduce the concrete concepts of atomic environment and the similarity function. In the section of Implementation, we discuss the issues on the practicality and functionality of this approach and show the cross-validation with the SQS method. In the section of Application, we take the typical quinary CoCrFeMnNi HEA, continuous solid-solution binary Ta-W alloy and ternary CoCrNi medium-entropy alloy with the SRO as the application prototypes for the SAE method. Combining with *ab initio* calculations, we investigate the structural properties and compare the calculation results with the available experiments. At the last section, some concluding remarks are presented.

## 2. Methodology

We start from two basic assumptions for a representative structure with fully chemical disorder. Assumption 1 is that the atoms were independently uniformly distributed in a certain sublattice. Assumption 2 is that there were similar environments of the atoms in a certain sublattice.

### 2.1 Basic concepts and notations

In this subsection, we introduce fundamental concepts and notations involved in the SAE method. First, an $m$-site cluster is formed by $m$ lattice sites in the supercell. Two $m$-site clusters are crystallographically equivalent if and only if their coordinates coincide under some lattice symmetry operations. These crystallographically equivalent $m$-site clusters constitute a class denoted by $A_m$. Specially speaking, the class of one-site cluster $A_1$ is referred as a sublattice. Secondly, the two $m$- site clusters are configurationally equivalent if and only if they coincide in both coordinates and occupation elements under certain lattice symmetry operations. These configurationally equivalent $m$-site clusters constitute a class denoted by $A_m^\sigma$ where $\sigma$ stands for a atomic configuration. In this paper, we tacitly assume that atomic configuration $\sigma$ is the distribution of chemical elements in the supercell and $A_m^\sigma$ comes from the subdivision of $A_m$ by $\sigma$. A schematic diagram of $A_m$ and $A_m^\sigma$ is illustrated in Fig.1. If the numbers of atom clusters in $A_m$ and $A_m^\sigma$ are denoted by respectively by $|A_m|$ and $|A_m^\sigma|$, then it holds

$$\sum_{A_m^\sigma} |A_m^\sigma| = |A_m|. \tag{2.1}$$

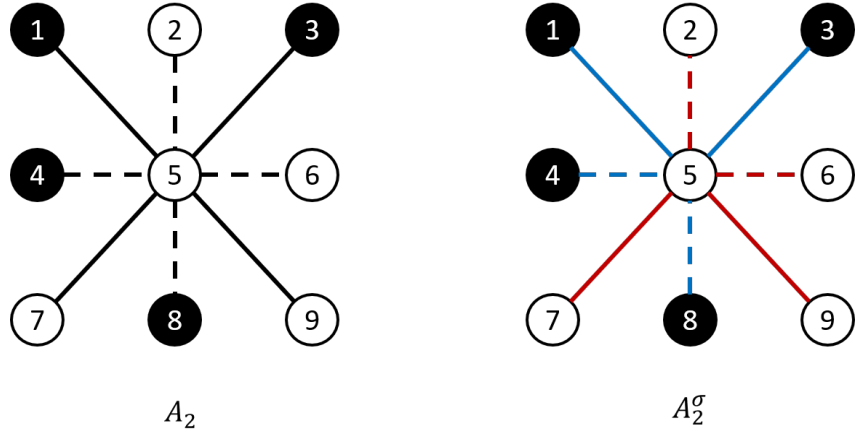

$A_2$ $\qquad\qquad\qquad A_2^\sigma$

Fig. 1. The 2-dim schematic diagram of atom clusters. There are two chemical elements represented by full and empty circles with the number index of the lattice site. In the left subfigure, the black dash and solid segments illustrate two types of crystallographically equivalent pair clusters. In the right subfigure, the blue and red solid segments illustrate two types of configurationally equivalent pair clusters subdivided from the cluster type represented by the black solid segments in the left subfigure. It is similar for the dash segments.

We consider the neighboring clusters related to a lattice site. A lattice site can be denoted by $\alpha_{i,j}$ which indicates the $j$-th site in the $i$-th sublattice. $A_m(\alpha_{i,j})$ is a class of crystallographically equivalent $m$-site clusters containing $\alpha_{i,j}$, and can be subdivided into configurationally equivalent types denoted by $A_m^\sigma(\alpha_{i,j})$. Likewise, $A_{m,i}$ is a class of crystallographically equivalent $m$-site clusters containing at least one site in the $i$-th sublattice. In a similar manner, we also define $A_{m,i}^\sigma$, $|A_m(\alpha_{ij})|$, $|A_m^\sigma(\alpha_{ij})|$, $|A_{m,i}|$ and $|A_{m,i}^\sigma|$.

## 2.2 Atomic environment and similarity function

For a lattice site $\alpha_{i,j}$, we define its individual atomic environment in terms of $A_m(\alpha_{i,j})$ as the following array

$$\vartheta[A_m(\alpha_{i,j}), \sigma] \equiv \left\{ \frac{|A_m^\sigma(\alpha_{ij})|}{|A_m(\alpha_{ij})|} \right\}_{A_m^\sigma(\alpha_{ij})}. \tag{2.2}$$

In an ordered structure, there is an identical atomic environment for each atom in certain sublattice. In a chemically disordered structure, the individual atomic environment could differ because of the random distribution of chemical elements in the sublattice. Actually, the statistical average of the individual atomic environments over random structures would be equal to each other, i.e.

$$\frac{E(|A_m^\sigma(\alpha_{i,1})|)}{|A_m(\alpha_{i,1})|} = \frac{E(|A_m^\sigma(\alpha_{i,2})|)}{|A_m(\alpha_{i,2})|} = \cdots\cdots = \frac{E(|A_m^\sigma(\alpha_{i,N_i})|)}{|A_m(\alpha_{i,N_i})|}, \quad (2.3)$$

where $E(\cdot)$ denotes the expectation and $N_i$ is the number of the lattice sites in the $i$-th sublattice. We regard Eq. (2.3) as the mathematical interpretation of the assumption (Assumption 2) that there were similar environments of the atoms in a certain sublattice. However, it could be difficult to simultaneously optimize each individual atomic environment to achieve Eq. (2.3) in a finite-size supercell. Instead, the overall atomic environment associated with the sublattice will be more suitable for finite-size supercell modeling. And the statistical average of the overall atomic environment can approximate well to that of the individual atomic environment when the supercell becomes sufficiently large, i.e.,

$$\frac{E(|A_m^\sigma(\alpha_{i,j})|)}{|A_m(\alpha_{i,j})|} = \frac{E(|A_{m,i}^\sigma|)}{|A_{m,i}|}, \quad j = 1, \cdots, N_i, \quad (2.4)$$

which is the underlying idea of the SAE method.

It is obvious from the crystallographical equivalence that

$$|A_{m,i}| = |A_m|, \quad |A_{m,i}^\sigma| = |A_m^\sigma|. \quad (2.5)$$

Then we define the overall atomic environment in terms of cluster type $A_m$ as follows

$$\vartheta(A_m, \sigma) \equiv \left\{\frac{|A_m^\sigma|}{|A_m|}\right\}_{A_m^\sigma}. \quad (2.6)$$

To calculate the expectation of the overall atomic environment, we consider a counting problem. Given one $m$-site cluster, its $m$ sites can be divided into $k$ sublattices and each one consists of $i_k$ lattice sites. We assign $n$ chemical elements to the $m$ sites. It is assumed that the $j$-th element is assigned to $d_{ij}(d_{ij} \geq 0, 1 \leq i \leq k; 1 \leq j \leq n)$ sites in the $i$-th sublattice. Among the atom clusters generated from all the above distributions, the number of configurationally equivalent $m$-site clusters of type $A_m^\sigma$ can be formulated as

$$\lambda(A_m^\sigma) = \frac{1}{N_1}\frac{i_1!}{d_{11}!d_{12}!\cdots d_{1n}!} \cdot \frac{1}{N_2}\frac{i_2!}{d_{21}!d_{22}!\cdots d_{2n}!} \cdots\cdots \frac{1}{N_k}\frac{i_k!}{d_{k1}!d_{k2}!\cdots d_{kn}!}, \quad (2.7)$$

where $N_i$ is a positive integer dependent on the symmetry of the $i$-th sublattice and the distribution of the chemical elements on the $m$-site cluster. For instance, $A_2^{pq}$ denote a type of pair cluster generated from the distribution of two elements $p$ and $q$. According to Eq. (2.7), $\lambda(A_2^{pq}) = \frac{1}{1} \cdot \frac{2!}{1! \cdot 1!} = 2$ if $A_2^{pq}$ coincides with $A_2^{qp}$ under some symmetry operation, or $\lambda(A_2^{pq}) = \frac{1}{2} \cdot \frac{2!}{1! \cdot 1!} = 1$ otherwise.

In the case of fully chemical disorder, the probability of finding the atom clusters of type $A_m^\sigma$ in all the clusters of type $A_m$ can be calculated as

$$P(A_m^\sigma|A_m) = \lambda(A_m^\sigma) \prod_{\substack{1\le i \le k \\ 1\le j \le n}} c_{ij}^{d_{ij}}, \qquad (2.8)$$

where the concentration of the $j$-th chemical element in the $i$-th sublattice is denoted by $c_{ij}(c_{ij} > 0)$. Eq. (2.8) is derived from the assumption (Assumption 1) that the atoms are independently uniformly distributed in the sublattice. The expectation of $|A_m^\sigma|$ is formulated as

$$E(|A_m^\sigma|) = P(A_m^\sigma|A_m) \cdot |A_m|. \qquad (2.9)$$

Now we define the similarity function associated with the atom clusters of type $A_m^\sigma$ as follows

$$f(A_m^\sigma) \equiv 1 - \frac{|A_m^\sigma|}{P(A_m^\sigma|A_m)|A_m|}, \qquad (2.10)$$

where $|A_m^\sigma|$ and $|A_m|$ can be counted and $P(A_m^\sigma|A_m)$ can be calculated as Eq. (2.8). This function can be used to measure the deviation of the current configuration from the fully disordered structure.

## 2.3 Generalized similarity function

In this subsection, we investigate the relationship between the similarity function as Eq. (2.10) and Warren-Cowley SRO parameter [28,29]. For simplicity, we assume that there is only one sublattice in the crystal. One way of defining the correlation function associated with $A_m$ is shown as

$$\gamma(A_m; p, q, \cdots, s) \equiv \langle c_{1;p} \cdot c_{2;q} \cdot \cdots \cdot c_{m;s} \rangle_{A_m} - c_p c_q \cdots c_s, \qquad (2.11)$$

where $p, q, \cdots, s$ represent chemical elements, $c_p, c_q, \cdots, c_s$ are corresponding concentrations in this sublattice, $c_{1;p}, c_{2;q}, \cdots, c_{m;s}$ are occupation numbers, and $\langle \cdots \rangle_{A_m}$ denotes the average over all crystallographically equivalent $m$-site clusters of type $A_m$. The occupation number $c_{1;p}$ takes on 1 if the component $p$ occupies the first site of the cluster, or zero otherwise. In a completely random structure, the occupation of the sites in any type of cluster is uncorrelated, i.e. $\gamma(A_m; p, q, \cdots, s) = 0$.

The most important correlation functions are pair functions, since they are usually the strongest ones and besides they can be measured in diffuse scattering experiments [30,31]. The Warren-Cowley SRO parameter can be defined as

$$\alpha_n^{pq} \equiv \alpha(A_{2(n)}; p, q) \equiv -\frac{\gamma(A_{2(n)}; p, q)}{c_p c_q}, \qquad (2.12)$$

where we use the coordination shell number $n$ to characterize the pair clusters of type $A_2$, and $p \ne q$ is usually assumed.

We denote by $A_{2(n)}^{pq}$ a type of the configurationally equivalent pair clusters which first and second sites are occupied by components $p$ and $q$, respectively. Simple derivation yields that

$$\begin{cases} \alpha_n^{pq} + \alpha_n^{qp} = 2f\left(A_{2(n)}^{pq}\right), & A_{2(n)}^{pq} = A_{2(n)}^{qp}; \\ \alpha_n^{pq} = f\left(A_{2(n)}^{pq}\right), & A_{2(n)}^{pq} \ne A_{2(n)}^{qp}. \end{cases} \qquad (2.13)$$

In a homogeneous solid-solution system, the pair correlation functions satisfy that $\alpha_n^{pq} = \alpha_n^{qp}$. In that case, we also have

$$\alpha_n^{pq} = f(A_{2(n)}^{pq}). \qquad (2.14)$$

Therefore, the similarity function can be regarded as a generalized SRO parameter for any type of cluster. In the case of pair cluster, we let $\theta\left(A_{2(n)}^{pq}\right) \equiv 1 - \alpha_n^{pq}$. More generally, we present the generalized similarity function as like

$$\tilde{f}(A_m^\sigma) \equiv \theta(A_m^\sigma) - \frac{|A_m^\sigma|}{P(A_m^\sigma|A_m)|A_m|}, \qquad (2.15)$$

where $\theta(A_m^\sigma)$ might be more or less than one to indicate the degree of SRO. As a result, the generalized similarity function could be used for a uniform quantitative description of fully chemical disorder and SRO.

## 2.4 Objective function of SAE method

To measure the deviation from the desired solid-solution structure, we establish the objective function of any type of cluster and current configuration $\sigma$ as follows

$$g(A_m, \sigma) \equiv \sqrt{\frac{1}{N(A_m^\sigma)} \sum_{A_m^\sigma} \tilde{f}^2(A_m^\sigma)} \qquad (2.16)$$

where $N(A_m^\sigma)$ denotes the number of the classes of configurationally equivalent cluster in $A_m$.

By the SAE method, to find a representative structure of the solid-solution system turns into minimizing the objective function of any type of clusters as Eq. (2.16) in the configuration space. In practice, it is likely sufficient to minimize the objective function for $A_2$ and $A_3$ clusters. And these implementation techniques will be discussed in Section 3.

It is worthwhile mentioning that the perfect configuration making Eq. (2.16) zero is also the optimal solution in perspective of the SQS method, i.e., the optimal solution in the SAE method provides a sufficient condition for the optimal solution in the SQS method. We will prove it in Appendix and illustrate by counterexample that the converse statement is not true.

# 3. Implementation

## 3.1 Practical objective function

Structural modeling by the SAE method starts from the selection of $A_m$ clusters considered in the objective function. It has been shown [27,32–34] that the alloy properties, such as formation enthalpy, lattice parameter and band gap, are indeed dominated by the interactions between near neighbors.

In practice, it is sufficient to minimize the objective function for $A_2$ and $A_3$ clusters to build a suitable configuration. The inclusion of longer ranged pairs and multibody clusters was attempted but did not yield any improvement in the cross-validation error [35]. For many applications, good accuracy can be obtained by retaining terms corresponding to only a relatively few pairs, three-body, and four-body clusters [36].

Furthermore, we define the size of the cluster as the maximum of the interatomic distances of any two sites in the cluster. The lattice clusters with smaller size are more important for the properties of materials. Thus, the two cutoff radii $r_2^c$ and $r_3^c$ are introduced for the cluster type $A_2$ and $A_3$ respectively. In order to make the smaller cluster contribute more in the objective function, we define the following weight function for the pair clusters:

$$w_2(A_2) \equiv \frac{\exp(-r/r^0)}{\sum_{r'} \exp(-r'/r^0)}, \tag{3.1}$$

where $r$ is the size of the two-site cluster of type $A_2$, $r^0$ is the minimum of the two-site cluster sizes, and $r'$ runs over all the sizes of the two-site cluster types. Likewise, we present the weight function for three-site clusters:

$$w_3(A_3) = \frac{\exp(-r_1/r_1^0)\exp(-r_2/r_2^0)\exp(-r_3/r_3^0)}{\sum_{r_1' r_2' r_3'} \exp(-r_1'/r_1^0)\exp(-r_2'/r_2^0)\exp(-r_3'/r_3^0)}, \tag{3.2}$$

where subscripts 1, 2 and 3 stand for the three interatomic distances of the three-site cluster, respectively. Then, we arrive at the practical objective function as follows

$$f(\sigma, r_2^c, r_3^c) = \sum_{d(A_2)<r_2^c} w_2(A_2)g(A_2, \sigma) + \sum_{d(A_3)<r_3^c} w_3(A_3)g(A_3, \sigma), \tag{3.3}$$

where $d(A_m)$ is the size of the cluster of type $A_m$.

In the ATAT [27] package, the objective function for the SQS method is presented as

$$Q(\sigma) = -wL + \sum_{A_m} \Delta\rho(A_m, \sigma) \tag{3.4}$$

where $w$ is a weight, $\Delta\rho(A_m, \sigma)$ is referred as Eq. (A.4), and $L$ is the largest $l$ such that $\Delta\rho(A_m, \sigma) = 0$ for all the clusters which size is not larger than $l$. To minimize Eq. (3.4), the disorder associated with small clusters should be optimized first. However, we can see that the objective function will have a sharp drop if a structure with a larger $l$ is found. We suggest that our weighted objective function could be a better choice, since small change in the configuration will not induce dramatic change in the objective function.

## 3.2 Efficient Metropolis Monte Carlo sampling

The configuration space grows exponentially with the number of atoms in the supercell, which makes it impossible to traverse all the configurations. We employ the Metropolis Monte Carlo (MMC) sampling to minimize the objective function algorithm. The MMC sampling method is also adopted by the ATAT package and has been demonstrated efficient in the structure prediction of atom cluster and crystal structure [30, 31]. Given a configuration $\sigma_s$ as a seed, the objective function is evaluated as $f(\sigma_s)$. A new configuration $\sigma_{new}$ is generated by randomly swapping the occupation chemical elements of two lattice sites in the same sublattice. Then update is continued by the Metropolis selection rule.

If $f(\sigma_{new}) < f(\sigma_s)$, the seed is updated to $\sigma_{new}$. Otherwise a random value $p$ is sampled in (0, 1) and compared with $e^{(f(\sigma_s)-f(\sigma_{new}))/(k_BT)}$, where $k_BT$ is a user-specified parameter to adjust the acceptance ratio. If $p < e^{(f(\sigma_s)-f(\sigma_{new}))/(k_BT)}$, the seed $\sigma_s$ is updated to $\sigma_{new}$. In other cases, the seed will not be updated. The procedure will be repeated for many times, until the objective function converges to a preset criterion, or the maximum number of structures has been achieved. Since the MMC algorithm is a relatively local method, and the optimized structure may be strongly influenced by the quality of the initial seed. In our implementation, a certain number of random configurations are generated and evaluated by the objective function. The most optimal random configuration is chosen as the seed for the MMC optimization. The evolution procedure should be repeated for several iterations.

In the current implementation of the SAE method, the two atoms to be swapped in the MMC step are randomly selected. If the atomic environment of each atom is analyzed, the atoms whose atomic environment is far from the fully disordered state should be chosen with higher priority.

Using the MMC algorithm, the SAE objective function is efficiently optimized. The desired finite size supercell to model the disordered solid solution is obtained when the objective function converged.

### 3.3 Support for 230 space groups

In practice, the symmetry and composition of the solid-solution materials may be more complicated than the bcc, fcc and hcp phases. In the implementation of the SAE method, we have created a database for Wyckoff positions of all 230 space groups. In this database, the multiplicity, symbol, coordinates for each Wyckoff position and the possible translation vector of the space group is recorded. With this database, the SAE method supports the modeling of multi-component and multi-sublattice solid-solution materials of any space group symmetry.

In some materials with multiple sublattices, such as high-entropy ceramics and perovskite, the atoms in the system are partially disordered. There might be a sublattice in the system, the lattice sites in which are occupied by just one type of element. The generation of partial disordered system is naturally supported by the SAE method. For the fixed sublattice, the ratio of the element could be set to 1, the required partial disordered structure could be generated using the SAE method.

The SAE method is also able to generate structures for the interstitial solid-solution. The multiple types of sublattices could be defined in the interstitial solid solution, one for the solvent atoms and the others for the interstitials. The interstitial lattice sites are occupied by the chemical atoms and vacancies. Once the sublattices and element concentrations are set, the structural modeling could be performed using the SAE method.

Furthermore, the input for the SAE method could also be greatly simplified. For example, in the $Im\bar{3}m$ (#229) space group, we suppose that there are two different sublattices in the crystal structure, Wyckoff positions *a* and *b*, and the total number of atoms in a unit cell is eight. If no space group symmetry is used, all coordinates of the eight atoms should be supplied as the input for SAE method. When the space group symmetry is used, only two representative coordinates for Wyckoff position *a* and *b* need to be supplied, and all other 6 coordinates could be generated by space group symmetry operations.

## 3.4 Treatment of the generalized SRO parameter

By introducing $\theta(A_m^\sigma)$ into the generalized similarity function as Eq. (2.15), we can optimize the configuration for the structure modeling of materials with SRO. However, it is difficult to determine the values of $\theta(A_m^\sigma)$ since they are mutually dependent. For instance, in a ternary ABC alloy, the values of $\theta(A_2^\sigma)$ with $A_2^\sigma$ being A-A cluster will also influence the value of $\theta(A_2^\sigma)$ with $A_2^\sigma$ being A-B cluster or B-B cluster, and may even influence the values of $\theta(A_3^\sigma)$ and so on. Here we present an alternative method to take SRO into the structure modeling by treating the SRO parameters as external constraints.

For each SRO parameter $\theta(A_m^\sigma)$, a minimum value $\theta_{min}(A_m^\sigma)$ and a maximum value $\theta_{max}(A_m^\sigma)$ are supplied. Using the SRO parameter $\theta(A_m^\sigma)$, the SRO related to the atom cluster composed of more than three atoms could also be considered. An SRO objective function is proposed as

$$\varepsilon(\sigma) = \sum_{\theta(A_m^\sigma)} \left| \frac{\theta_{min}(A_m^\sigma) + \theta_{max}(A_m^\sigma)}{2} - \frac{|A_m^\sigma|}{P(A_m^\sigma|A_m)|A_m|} \right|, \qquad (3.5)$$

which sum over all cluster with SRO. This function $\varepsilon(\sigma)$ could also be effectively optimized by the MMC algorithm until all the SRO parameters fall into the given ranges.

Once the structure that fulfills all SRO constraints is generated, the structure will be further optimized according to the SAE objective function as Eq. (2.15). With SRO constraints, the SRO parameters will be checked in each MMC step. Only structures that fulfill the SRO constraints will be chosen as seeds in the optimization procedure. In this approach, we are able to get the most disordered configurations that fulfill the external SRO constraints.

Figure 2 shows the flowchart of the SAE structure generation. The part of the MMC algorithm with constraints is marked in red, which is further illustrated in Fig. 3. Note that for the SAE generated configurations with SRO constraints, the validated structure should be further determined by the free energy calculation or experiment observation.

In all, depending on the local atomic environments in real alloys, we can apply the SAE method to generate the solid-solution structure with fully disorder, partial disorder, short range order as well as the solid-solution structures with interstitial atoms.

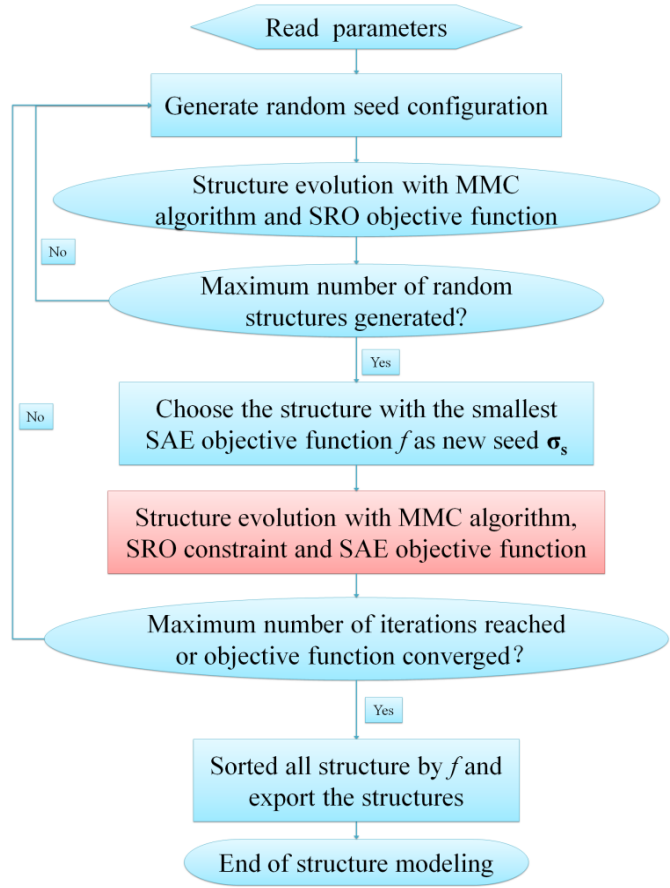

Fig. 2. Flowchart of the SAE method.

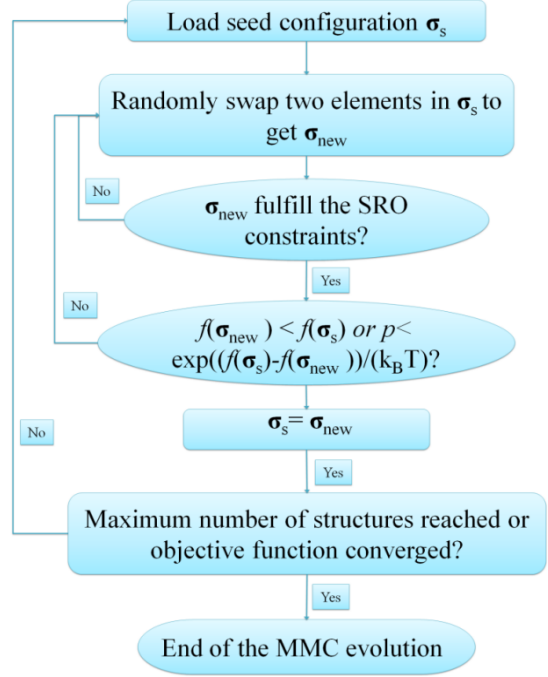

Fig. 3. Flowchart of the MMC algorithm with SRO constraints.

## 3.5 Cross validation of SAE and SQS methods

Cross-validation between the SQS and SAE methods is performed using six types of solid-solution structures, *i.e.* equimolar binary AB alloy and ternary ABC alloy for bcc, fcc and hcp phases. Structural models are generated using both SQS and SAE methods. We consider the two-site cluster $A_2$ and three-site cluster $A_3$ in both methods. The mcsqs program in the ATAT package is used as the implementation of SQS method [27].

For each type of solid-solution structures, we employed the same supercell so as to compare the SAE and SQS methods at the same level. The lattice constants of the unit cells and the sizes of the supercells are listed in Table 1.

Table 1 The lattice constant (in Å), size of supercell of the bcc, fcc and hcp. The $R_1$ and $R_2$ represent the first and second nearest neighboring distances.

| Structure | Lattice constants | Supercell | #Atoms | $R_1$, $R_2$ |
|---|---|---|---|---|
| bcc | $a=b=c=1.0$ | 3x3x3 | 54 | $\sqrt{3}/2$, 1.0 |
| fcc | $a=b=c=1.0$ | 3x3x3 | 108 | $\sqrt{2}/2$, 1.0 |
| hcp | $a=b=1.0$, $c=1.63299$ | 3x3x3 | 54 | 1.0, $\sqrt{2}$ |

For both SAE and SQS methods, the MMC algorithm is applied to optimize the objective functions. We limited the maximum number of configurations to be 100,000 in each iteration for both methods. For each combination of composition and cutoff radii, the calculation will be repeated for three times, and only the best structures in the three iterations are reported here. For the SAE method, the 100,000 configurations are divided to 10 sub-iterations. 1000 random configurations are generated and evaluated in each sub-iteration, and the configuration with the smallest objective function is chosen as the seed configuration for the MMC stage. The rest 9000 configurations are generated in the MMC stage. It is not guaranteed that the optimal structure could be obtained within limited MMC steps, but the results are still useful.

Table 2 Cross-validation between the SAE method and the SQS method for the typical binary and ternary alloys. The deviations from the desired configuration in terms of two-site cluster and three-site cluster are measured by the SAE criterion and SQS criterion, respectively. In the SAE method, the deviation is measured by the objective function as Eq. (3.3); in the SQS method, it is measured by the discrepancy in the two-site or three-site correlation function. These deviations are denoted by $\Delta_2^{SAE}$, $\Delta_3^{SAE}$, $\Delta_2^{SQS}$ and $\Delta_3^{SAE}$, respectively.

| Symmetry | Composition | Cutoff (Å) | SAE Structure | | | | SQS Structure | | | |
|---|---|---|---|---|---|---|---|---|---|---|
| | | | $\Delta_2^{SAE}$ | $\Delta_3^{SAE}$ | $\Delta_2^{SQS}$ | $\Delta_3^{SQS}$ | $\Delta_2^{SAE}$ | $\Delta_3^{SAE}$ | $\Delta_2^{SQS}$ | $\Delta_3^{SQS}$ |
| bcc | AB | 0.9 | 0.000 | 0.000 | 0.000 | 0.000 | 0.000 | 0.000 | 0.000 | 0.000 |
| | | 1.05 | 0.006 | 0.019 | 0.012 | 0.000 | 0.006 | 0.019 | 0.012 | 0.000 |
| | ABC | 0.9 | 0.000 | 0.000 | 0.000 | 0.000 | 0.000 | 0.000 | 0.000 | 0.000 |
| | | 1.05 | 0.000 | 0.000 | 0.000 | 0.000 | 0.000 | 0.169 | 0.000 | 0.024 |
| fcc | AB | 0.75 | 0.000 | 0.000 | 0.000 | 0.000 | 0.000 | 0.000 | 0.000 | 0.000 |
| | | 1.05 | 0.000 | 0.005 | 0.000 | 0.000 | 0.000 | 0.005 | 0.000 | 0.000 |
| | ABC | 0.75 | 0.000 | 0.000 | 0.000 | 0.000 | 0.000 | 0.008 | 0.000 | 0.003 |
| | | 1.05 | 0.000 | 0.007 | 0.000 | 0.013 | 0.011 | 0.063 | 0.017 | 0.046 |
| hcp | AB  | 1.05 | 0.012 | 0.021 | 0.025 | 0.000 | 0.012 | 0.021 | 0.025 | 0.000 |
|     |     | 1.50 | 0.012 | 0.042 | 0.037 | 0.000 | 0.012 | 0.099 | 0.037 | 0.000 |
|     | ABC | 1.05 | 0.000 | 0.007 | 0.000 | 0.072 | 0.000 | 0.181 | 0.000 | 0.265 |
|     |     | 1.50 | 0.000 | 0.105 | 0.000 | 0.130 | 0.025 | 0.285 | 0.072 | 0.297 |

The optimal configurations generated by SQS and SAE methods are then further crossly evaluated by the objective function of SAE and the summation of deviations of the SQS methods, respectively. The deviations of the optimal configurations generated by using the SAE and SQS methods are listed in Table 2.

For the bcc structures, when the cutoff radii are both 0.90 Å, only one type of $A_2$ cluster is considered while no $A_3$ clusters are considered since even the smallest $A_3$ cluster has an interatomic distance of 1.0 Å. Under such circumstance, both SAE and SQS methods are able to generate the optimal configurations with objective functions being zero.

For the fcc AB and ABC structures with cutoff radii being 0.75 Å, one $A_2$ cluster and one $A_3$ cluster are considered in the both methods. The SAE method is also able to generate optimal configuration with zero objective functions for AB and ABC alloys. Meanwhile, the SQS method failed to generate the optimal configuration for ternary ABC alloy, due to the limited number of configurations or unreasonable seed configuration in the test runs.

When the cutoff radii for $A_2$ and $A_3$ clusters are increased to 1.05 Å, more lattice clusters are included in the objective functions of two methods. For the bcc binary AB alloy, the SAE structure has the same quality with the SQS structure. Interestingly, though both structures are perfect configurations evaluated by the SQS criterion in terms of three-site cluster, i.e., $\Delta_3^{SQS} = 0$, neither is the optimal solution in terms of three-site cluster under the SAE criterion. This is just the counterexample we have mentioned in Section II.4. For the bcc ternary ABC alloy, the SAE method successfully generates the perfect configuration in both the SAE and SQS criteria, while the SQS method generates the configuration which $\Delta_3^{SAE}$ and $\Delta_3^{SQS}$ are nonzero.

For the fcc ternary ABC alloy, the SAE method is able to generate configurations with $\Delta_2^{SAE} = 0$ and a small $\Delta_3^{SAE}$. Meanwhile, the configuration produced by the SQS method also has small $\Delta_2^{SAE}$ and $\Delta_3^{SAE}$, but these values are larger than those of the SAE configurations. For the fcc AB alloy, we also notice that the $\Delta_3^{SQS}$ values of both configurations vanish, but their values of $\Delta_3^{SAE}$ do not. This is another counterexample illustrating that a perfect SQS configuration is not sufficient for the optimal solution of the SAE method.

For the hcp AB and ABC structures, the SAE configuration has the same or better quality compared with the SQS configuration. For the AB structures with the cutoff radii being 1.05 Å, they are equally good. When the cutoff radius increases to 1.50 Å, more clusters are included in the objective function, and the quality of the SAE structure is better than that of the SQS structures. For hcp ABC structures, it holds that $\Delta_2^{SQS} = 0$ when $\Delta_2^{SAE} = 0$. In the contrary, for the hcp AB structures, $\Delta_3^{SAE} \neq 0$ when $\Delta_3^{SQS} = 0$.

Although different objective functions are adopted, the SAE method is comparable with the SQS method. The cross comparison suggested that any optimal configuration generated by SAE is sufficiently optimal for SQS.

Moreover, we performed simulation on the MoNbTa alloy and compared with previous simulation using the SQS method. The equimolar MoNbTa alloy is modeled with SAE method using supercells with 24, 36, 54 and 72 atoms. The lattice constants and mixing enthalpies per atom are calculated and compared with previous SQS work [34]. The *ab initio* calculation details are the same to the SQS work. The calculation results are shown in Table 3.

Table 3 The average lattice constants *a* (in Å) and mixing enthapies Δ*H* (eV/atom) of MoNbTa equimolar alloy with supercell of different sizes (the number of atoms *N*).

| N  | a      | ΔH      |
|----|--------|---------|
| 24 | 3.2549 | -0.0726 |
| 36 | 3.2544 | -0.0761 |
| 54 | 3.2544 | -0.0770 |
| 72 | 3.2546 | -0.0768 |

From the calculation results, we can see that the lattice constants converge well at 24 atoms. The mixing enthalpy of the 24-atom supercell is within 0.01 eV/atom with larger supercells. Our calculation results also agrees well with previous SQS work.

## 4. Application

### 4.1 *Ab initio* calculation details

Based on the SAE methods, we have constructed supercell models for several typical solid-solution alloys and performed *ab initio* calculations to investigate their phase stabilities, so as to demonstrate the validity of SAE method.

In the present *ab initio* calculations, we employed the Vienna *Ab initio* Simulation Package (VASP) computer program [32] based on the density functional theory [8,9]. The exchange-correlation potentials were treated by the Perdew-Burkey Ernzerhof [33] functional within the generalized gradient approximation. The electron-ion interaction was described by the projector augmented wave (PAW) method [34]. The plane-wave cutoff energy is 300 eV. The Brillouin zone sampling was performed using the special *k* points generated by Monkhorst-Pack scheme [35] with density parameters 0.2 Å$^{-1}$. The convergence tolerance level is 10$^{-6}$ eV for total energy and 0.01eV/Å for the max force on each atom, respectively.

In the calculations of ferromagnetic Co-Cr-Fe-Mn-Ni alloys, opposite spin moments are set for Cr, Mn atoms and Fe, Co, Ni atoms, *i.e.*, Cr and Mn are set as spin up, while Fe, Ni and Co are set as spin down. The paramagnetic state above the Curie temperature is modeled by using the disordered local moment (DLM) approximation [36]. Namely, the spin-up and spin-down atoms with equal atomic fraction for the same elements are treated as different atomic species distributed randomly in the supercell. For example, for a 180-atom CoCrFeMnNi quinary equimolar alloy, the SAE structure with PM state contains 10 components (18Co↑18Co↓)(18Cr↑18Cr↓)(18Fe↑18Fe↓)(18Mn↑18Mn↓)(18Ni↑18Ni↓).

## 4.2 Phase stability of CoCrFeMnNi alloy

As a typical HEA sample, single phase CoCrFeMnNi [37] alloy has drawn much attention since being reported in 2004. The experimental lattice parameter of fcc structure is about 3.590~3.610 Å, the corresponding equilibrium volume is 11.56~11.76 Å$^3$/atom. As far as we know, the experimental CoCrFeMnNi HEA was defaulted as the paramagnetic (PM) state, due to the heat treatment. The chemical elements adopt different magnetic states in their ground state: ferromagnetic (FM) for Fe, Ni, Co, and antiferromagnetic for Cr, and multi-magnetic for Mn. In the following section, the FM state represents the ferromagnetic set for Fe, Ni, Co, Mn and antiferromagnetic set for Cr, while AM state stands for antiferromagnetic set for Cr, Mn and ferromagnetic set for Fe, Ni, and Co in our *ab initio* calculations. For the PM state, the DLM approximation was employed. For the no magnetic (NM) state, we considered non spin polarized set in *ab initio* calculations.

Using the SAE method, we constructed a 180-atom supercell to model the disordered solid-solution structure, *i.e.* 3×3×5 fcc SAE structure (see Fig.4 A). The equilibrium volume predicted by *ab initio* calculations is 11.07 Å$^3$/atom for FM state, 11.06 Å$^3$/atom for PM state, and 11.04 Å$^3$/atom for AM state, respectively. They are slightly smaller than the experimental results, but agree well with the available results from the CPA calculations (10.98 Å$^3$/atom [38] and 11 Å$^3$/atom [39]). Recent *ab initio* calculations suggested that hcp is more stable than fcc at *T*=0 K [38,39], which agrees with more recent experiments [40,41]. To estimate the validity of SAE method for the phase stability, we constructed the hcp based SAE structure, *i.e.* 5×5×3 hcp supercell with 150 atoms (see Fig.4 B). From Table 4, we can find that at each magnetic state, the 150-atom hcp SAE structure is indeed more stable than the 180-atom fcc SAE structure in the ground state.

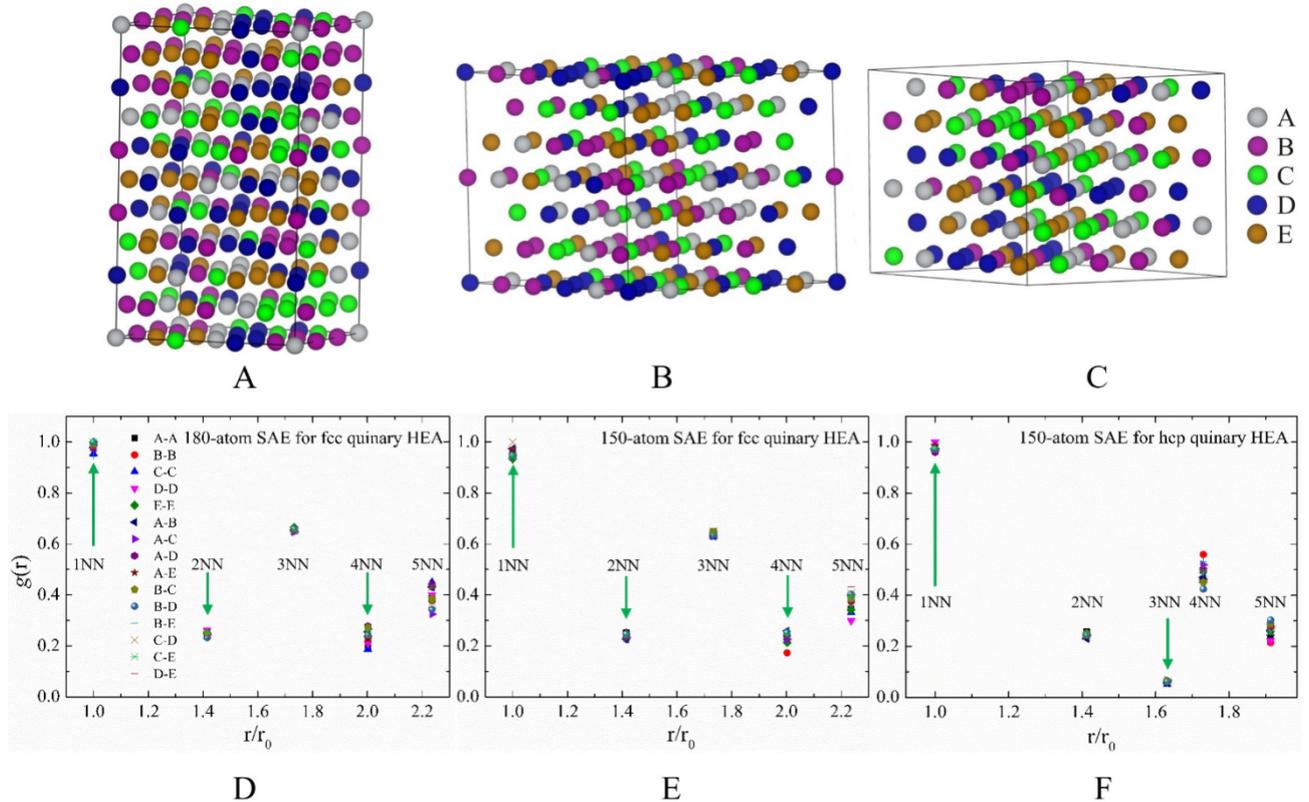

Fig. 4. SAE structures (A, B, C) and the corresponding radial distribution function g(r) (D, E, F) as a function of the 1NN-5NN atom pairs. $r_0$ is the first neighboring atom distance. A and D for 180-atom fcc SAE structure, B and E for 150-atom fcc SAE structure based on ABC stacked configuration implemented along the fcc <111> crystal direction, and C and E for 150-atom hcp SAE structure.

Table 4 Equilibrium volume per atom $V$ (Å$^3$), energy per atom $E$ (eV/atom), and the magnetic moment per atom $\mu$ ($\mu_b$) for the different magnetic states (ferromagnetic FM, antiferromagnetic AM, paramagnetic PM, and no magnetic (without spin polarized set) NM) of fcc and hcp CoCrFeMnNi.

| HEA | $V$ (180) fcc | $E$ | $\mu$ | $V$ (150) fcc <111> | $E$ | $\mu$ | $V$ (150) (hcp) | $E$ | $\mu$ |
|---|---|---|---|---|---|---|---|---|---|
| FM | 11.07 | -7.781 | 0.58 | 10.07 | -7.781 | 0.51 | 10.93 | -7.802 | 0.22 |
| AM | 11.06 | -7.783 | 0.37 | 10.07 | -7.783 | 0.41 | 10.90 | -7.794 | 0.15 |
| PM | 11.04 | -7.780 | 0.03 | 11.01 | -7.781 | 0.07 | 10.87 | -7.791 | 0.09 |
| NM | 10.69 | -7.751 | 0 | 10.69 | -7.754 | 0 | 10.67 | -7.786 | 0 |

In order to eliminate error caused by size effect, we then constructed the SAE structure with the same atom number to model the hcp and fcc CoCrFeMnNi HEA. Along the fcc<111> crystallographic direction, we constructed ABC stacked hexagonal structure and enlarged the 3-atom fcc<111> hexagonal structure to a 150-atom fcc<111> SAE structure, *i.e.* 5×5×3 supercell (see Fig.4 C). From Table 4, we can see that the 150-atom hcp SAE structure is still more stable than the 150-atom fcc <111> SAE structure. Interestingly, both *ab initio* predicted average energies per atom and the equilibrium volumes of fcc SAE and fcc<111> SAE structures are excellently consistent with each other for different magnetic states. For fcc SAE and fcc <111> SAE structures, the total magnetic moment per atom at the FM state is very close to that of the AM state. Whereas the magnetic moment of PM state is close to zero, which may suggest the validity of DLM approximation application to the SAE method.

Figure 4 (D, E, F) shows the radial distribution function of the three SAE structures above. The quinary alloy has 15 different types of atom pairs. Due to the equimolar ratio of chemical elements, the radial distributions are very close to each other for the 1th to 5th nearest neighboring atom pairs.

### 4.3 Partial disorder of Ta-W alloy

From the fully ordered phase to the fully disordered phase, there often exists the partial disordered phase. For an equimolar binary alloy with B2 structure, it has two different sublattices (*A* and *B*). Taking binary Ta-W alloy as an example, we use the concentration of W (*x*) located on the *A* lattice site to define the partial disorder in the Ta-W alloy. When *x*=0, Ta (W) occupies the A (B) lattice sites, the Ta-W alloy form a fully ordered B2 phase. Whereas *x*=0.5, Ta and W are uniformly distributed on A and B lattice sites, and the Ta-W alloy forms a random solid solution with bcc crystal structure. The partial disorder is corresponding to the value *x* (0<*x*<0.5).

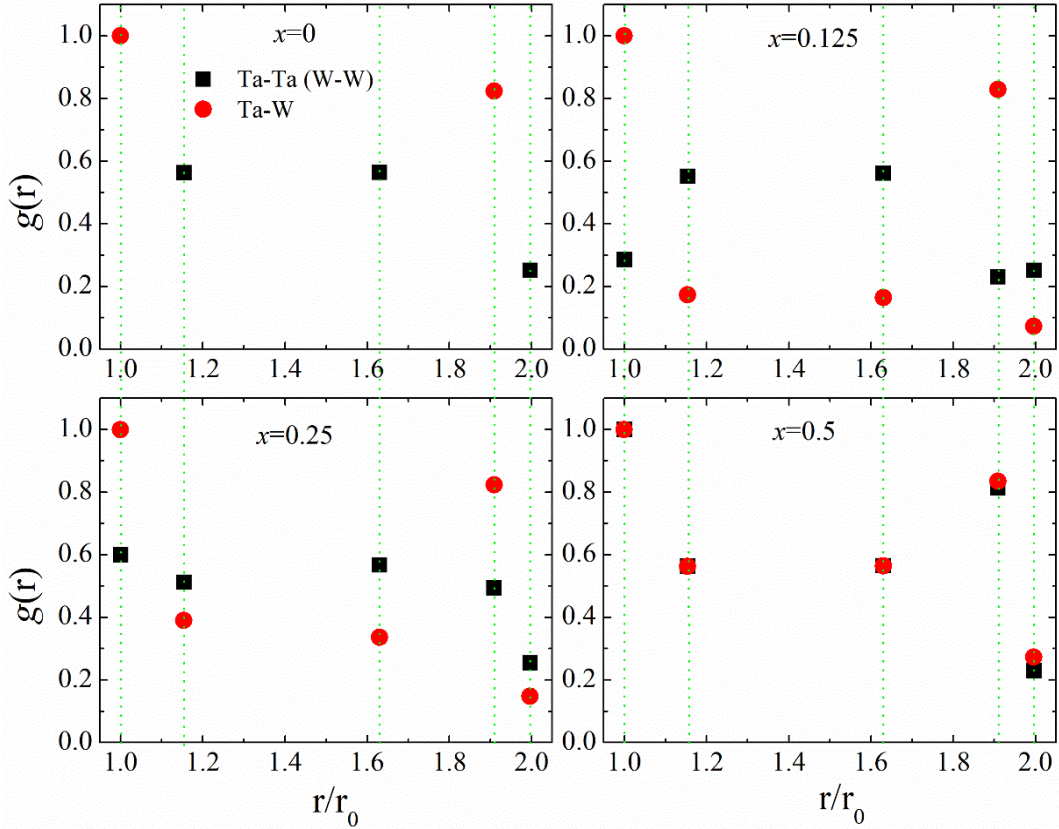

Fig. 5. Radial distribution $g(r)$ as a function of the nearest neighboring atom pairs $N$=1-5 for the 54-atom SAE Ta-W alloys from fully order B2 ($x$=0) to partial disordered ($x$=0.125, 0.25) to fully disordered bcc ($x$=0.5). $r_0$ is the first nearest neighboring atom distance.

Figure 5 shows the radial distribution $g(r)$ as a function of the nearest neighboring (NN) atom pairs. The $g(r)$ of the first nearest neighboring (1NN) atom pairs is normalized as 1. The 1NN atom number is set as $n_0$. We define the $N$th nearest neighboring atom distance and atom number as $r$ and $n$, respectively. The $g(r)$ of the $N$th nearest neighboring atom pairs satisfies $g(r) = (n/n_0)\times(r_0/r)^2$. For example, in the bcc crystal structure with lattice parameter $a_0$, with respect to the 1NN atom distance, the 2NN and 3NN atom distances should be equal to ($\sqrt{3} a_0/2)/a_0 \sim 1/1.155$ and $(\sqrt{3}a_0/2)/(\sqrt{2}a_0)\sim 1/1.633$, respectively. When $x$=0, Ta and W form the fully ordered B2 structure. The 1NN atom pairs are all Ta-W. The number of 1NN atom pairs should be 8. The 2NN atom pairs are Ta-Ta or W-W. The number of the 2NN (3NN) atom pairs is 6 (12). When the 1NN atom number is normalized, the radial distribution of the 2NN (3NN) atom pair is equal to $g(r)$ = $(6/8)\times(1/1.155)^2$=0.562 $((12/8)\times(1/1.633)^2$=0.563).

From Fig. 5, we see that the atom pair (W-W, Ta-W or Ta-Ta) orderly occupies on the nearest neighboring, whereas the Ta-Ta, W-W and Ta-W atom pairs exist in the bcc phase and they have similar radial distribution from 1NN to 5NN atom pairs. For the partial disordered phase, the radial distribution of Ta-W atom pair is not equal to that of Ta-Ta and W-W atom pairs. The number of atom pairs depends on the different nearest neighboring environments. With increase of W concentration $x$, the number of Ta-W atom pairs become close to Ta-Ta (W-W) atom pairs. Table 5 lists the energy difference of Ta-W alloys. The energy difference becomes larger with increase of disorder degree (from the fully ordered B2 to fully disordered bcc). It suggests that the stable phase of Ta-W alloy tends to form the ordered B2 structure at $T = 0$ K.

Table 5 Listed are the energy differences $\Delta E$ (meV/atom) of Ta-W binary alloy, the partial disorder is represented by the content $x$ (0, 0.125, 0.25, 0.5) of W occupying on A lattice in B2 crystal structure. The reference energy is that of B2 Ta-W alloy.

| Ta-W | Sub lattice (A, B) | $\Delta E$ |
|---|---|---|
| B2 | A-Ta1.00W0.0, B-Ta0.0W1.0 | 0.00 |
| ↓ | A-Ta0.875,W0.125, B-Ta0.125,W0.875 | 9.73 |
|  | A-Ta0.75,W0.25, B-Ta0.25,W0.75 | 16.18 |
| bcc | A, B-Ta0.5,W0.5 | 21.77 |

## 4.4 Short-range order in CoCrNi alloys

Firstly, the stability of CoCrNi without SRO is studied. To eliminate the error caused by size effect on the phase stability, we constructed a 54-atom SAE configuration for fcc CoCrNi solid solution [42], *i.e.*, 3×3×3 hcp supercell for hcp solid solution and 3×3×2 fcc<111> supercell for fcc solid solution. Fig. 6 shows the energy of hcp and fcc solid solution with different magnetic states. We can find that the hcp phase is more stable than fcc at different magnetic states. In fact, when CoCrNi is considered as a fully disordered solid solution, the calculation results based on CPA and SQS all suggested that fcc is not the ground state structure, with respect to the double hexagonal close packed (dhcp) and hcp phases [43]. Considering that the size of SAE structure for CoCrNi is slightly small, we constructed three SAE configurations for hcp and fcc phases, respectively. Note that the fcc SAE structures are based on the fcc <111> configuration (ABC stacked hexagonal structure). For the hcp or fcc SAE structures, the difference of energy per atom between three SAE configurations is smaller than $5 \times 10^{-3}$ eV/atom for the same magnetic state.

Secondly, the stability of CoCrNi with SRO is studied. Recent experiments indicated that there is SRO near Cr atoms (close to (Ni,Co)$_2$Cr) in the fcc CoCrNi alloy [44]. In our SAE structural model, the new SRO parameter is introduced as a constraint for the consideration of the SRO. Fig. 7 shows the illustrated distribution of the first nearest neighboring atom pairs near one Cr atom. For the fully disordered CoCrNi alloy, the SRO parameter $\theta$ should be equal to 1 for each atom pair. For the CoCrNi with SRO (near Cr), there are less Cr atoms in the first nearest neighbors of Cr atoms, so the value of $\theta_2^1(\mathrm{Cr},\mathrm{Cr})$ should be smaller than 1. During the structural modeling, the minimum and maximum values for the SRO parameter $\theta_2^1(\mathrm{Cr},\mathrm{Cr})$ are 0.75 and 0.825 respectively. Fig. 8 shows the radial distribution of the nearest neighboring atom pairs for the CoCrNi solid solution with SRO. For the 1NN atom pairs, the number of Cr-Ni and Cr-Co is 88 and 92, respectively, whereas the number of Cr-Cr is 60. The SRO in the model structure is obvious. Meanwhile, similar neighboring distribution is kept for the 2NN~5NN atom pairs.

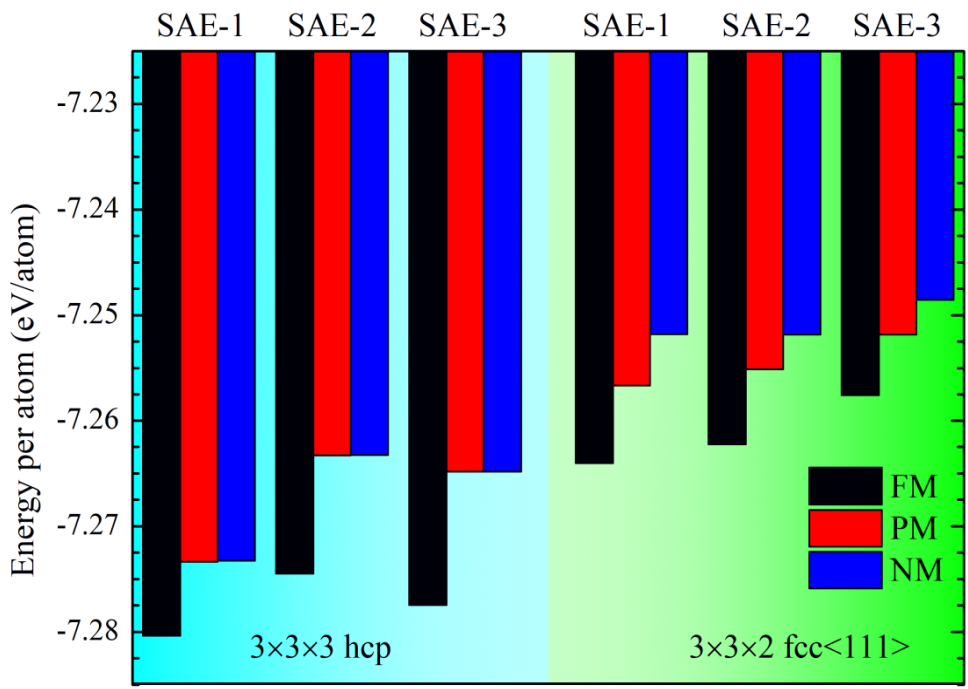

Fig. 6. Average energy per atom $E$ (eV/atom) of the random CoCrNi solid solution at different magnetic states. SAE-1, SAE-2 and SAE-3 represent the three SAE configurations with 54 atoms for hcp and for fcc<111> phases. FM, PM and NM represent the ferromagnetic, paramagnetic and non magnetic states, respectively.

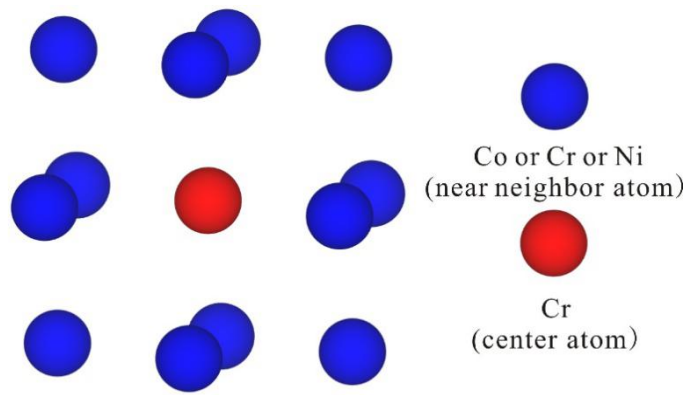

Fig.7. Illustrative configuration of the CoCrNi alloy with SRO.

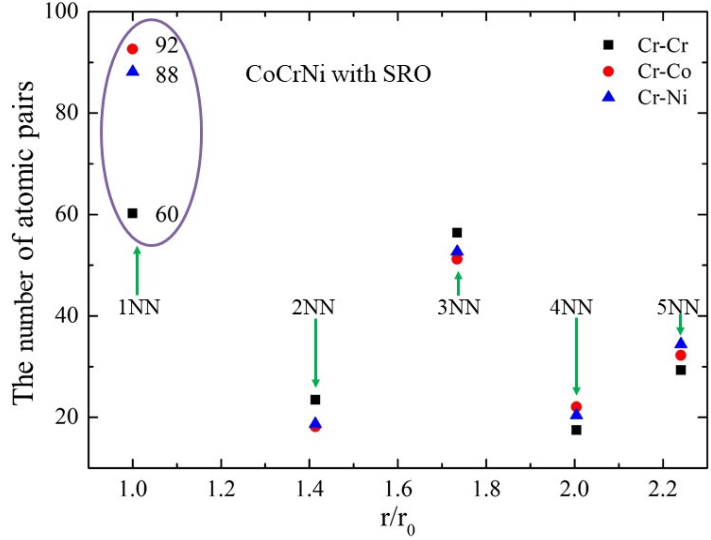

Fig. 8. The number of the *N*th (*N*=1-5) nearest neighboring (NN) atom pairs for CoCrNi with SRO. $r_0$ is the first neighboring atom distance.

Table 6 Equilibrium volume per atom *V* (Å$^3$), average energy per atom *E* (eV/atom), and magnetic moment per atom *μ* ($μ_B$) for SAE structures for CoCrNi with/without SRO, respectively. There are 108 atoms in each SAE structure. FM, PM and NM represent the ferromagnetic, paramagnetic and non-magnetic states, respectively.

|  | fcc SAE with SRO | | | fcc SAE without SRO | | |
|---|---|---|---|---|---|---|
| SAE | *V* | *E* | *μ* | *V* | *E* | *μ* |
| 1(FM) | 10.89 | -7.27308 | 0.380 | 10.96 | -7.25780 | 0.516 |
| 1(PM) | 10.88 | -7.27222 | 0.090 | 10.94 | -7.25507 | 0.226 |
| 1(NM) | 10.82 | -7.27003 | 0 | 10.85 | -7.24931 | 0 |
| 2(FM) | 10.91 | -2.27664 | 0.431 | 10.95 | -7.25955 | 0.642 |
| 2(PM) | 10.89 | -7.27458 | 0.01 | 10.94 | -7.25768 | 0.357 |
| 2(NM) | 10.82 | -7.27033 | 0 | 10.85 | -7.25044 | 0 |
| 3(FM) | 10.90 | -7.27722 | 0.334 | 10.95 | -7.26018 | 0.596 |
| 3(PM) | 10.89 | -7.27498 | 0.01 | 10.92 | -7.25783 | 0 |
| 3(NM) | 10.81 | -7.27259 | 0 | 10.84 | -7.25263 | 0 |

Table 6 shows the equilibrium bulk properties of the CoCrNi solid solution with/without SRO. Results suggest that the SRO makes the fcc structure more stable than the hcp phase. Due to the existence of antiferromagnetic Cr atoms, the Curie temperature of CoCrNi is very low (<5K) [19]. According to the available estimation of Curie temperature, our *ab initio* predicted Curie temperature is about 6.8~21.2 K from three different SAE configurations. For the CoCrNi solid solution with SRO, three SAE configurations have similar average magnetic moment close to 0.4 $μ_B$ at ferromagnetic state, while the average magnetic moment of paramagnetic state is close to zero. Whereas the average magnetic moment of random CoCrNi solid solution is slightly large, with respect to the CoCrNi solid solution with SRO, *i.e.,* SRO induces the decrease of magnetic order. In the three magnetic states, the ferromagnetic state has the largest equilibrium volume. The increasing equilibrium volume of CoCrNi without SRO may derive from the magnetic order.

# 5. Conclusion

We have presented a new method for the description of atomic environment (AE) of an atom and introduced the similar atomic environment (SAE) approach for structure modeling of the solid-solution materials. The SAE approach provides a uniform way to take into account both the full disorder and short-range order via minimizing the objective function in the configuration space. A new type of short-range order parameters based on the concept of atom cluster is introduced, which can provide finer characterization of SRO in solid solution. The SAE method could be applied to analyze the effect of different SRO on the stability of solid solution, providing more details in the structures of real solid solution.

We prove theoretically that the optimal structure generated by the SAE method must be the optimal structure of the special quasi-random structure (SQS) method and find several numerical counterexamples to the necessity. Based on the delicate implementation, an effective method based on Metropolis Monte Carlo algorithm is provided for the optimization of the objective function. We are enabled to extend the applicability of SAE method from full disorder to partial disorder and from substitutional solid solution to interstitial one.

It is demonstrated by typical alloy applications that the supercell configurations generated by the SAE method are able to describe the complicated solid solutions. For the CoCrFeMnNi high-entropy alloy composed of multiple magnetic elements, the ferromagnetic, antiferromagnetic and paramagnetic states have very close equilibrium volume and energy, although their average magnetic moments are slightly different. An *ab initio* calculation on Ta-W continuous solid-solution binary alloy with partial disorder suggested that Ta-W has the tendency to form ordered B2 phase at 0K. The SRO plays a key role in the phase stability and magnetic order in the CoCrNi medium-entropy alloy [44].

In general, the SAE method allows investigating the phase transformation from the fully disordered to short-range ordered solid solution based on the free energy calculations. Recently, the SAE method has been used to characterize the chemical environment effects on point defect formations in CoCrNi-based concentrated solid-solution alloys [52]. We conclude that the combination of the SAE approach with *ab initio* calculations offers an effective approach to the study on the phase stability, mechanical properties and electronic structure for solid-solution materials.

**Acknowledgements**

This work was supported by the Science Challenge Project (Grant No. TZ2018002) and the National Key Research and Development Program of China (Grant No. 2016YFB0201204). Authors also acknowledge the National Science Foundation of China (Grant Nos. 51701015, 91730302, 11501039, U1804123).

# Appendix

The SQS method is one of the most popular supercell methods applied in structural modeling for the chemical disordered materials [27]. We feel obliged to clarify the relationship between the SAE and SQS methods.

**Proposition**: For the completely random structure, the perfect configuration of the SAE method must be an optimal solution for the SQS method.

Our proof is as follows:

For a lattice cluster $A_m$, the correlation function defined in the SQS method is formulated as

$$\rho(A_m, \sigma) = \langle \Gamma(a_m^\sigma) \rangle_{A_m} = \frac{\sum_{a_m^\sigma} \Gamma(a_m^\sigma)}{|A_m|} \quad (A.1)$$

where $a_m^\sigma$ stands for an $m$-site cluster with the given configuration $\sigma$, and $\Gamma(a_m^\sigma)$ is the corresponding cluster function. The summation in Eq. (A.1) runs over all possible clusters in the supercell. Since the configurationally equivalent atom clusters share the same cluster functions, Eq. (A.1) can be rewritten

as

$$\rho(A_m, \sigma) = \sum_{A_m^\sigma} \frac{|A_m^\sigma|}{|A_m|} \Gamma(A_m^\sigma) \quad (A.2)$$

From the above formula, it can be observed that the correlation function is solely determined by the numbers of the clusters of type $A_m^\sigma$ in the supercell. For the fully disordered structure, the atoms in the sublattice are independently and uniformly distributed. So, the correlation function can be calculated as

$$\rho(A_m, \sigma^*) = \sum_{A_m^\sigma} P(A_m^\sigma | A_m) \cdot \Gamma(A_m^\sigma), \quad (A.3)$$

where $\sigma^*$ stands for the completely random structure, and the conditional probability $P(A_m^\sigma | A_m)$ is calculated as Eq. (2.8).

The SQS deviation function between a finite supercell configuration and fully disordered solid solution is defined as

$$\Delta\rho(A_m, \sigma) \equiv \rho(A_m, \sigma) - \rho(A_m, \sigma^*) = \sum_{A_m^\sigma} \left[\frac{|A_m^\sigma|}{|A_m|} - P(A_m^\sigma | A_m)\right] \cdot \Gamma(A_m^\sigma). \quad (A.4)$$

The perfect configuration of the SAE method makes the corresponding similarity function as Eq. (2.10) vanish, i.e.,

$$1 - \frac{|A_m^\sigma|}{P(A_m^\sigma | A_m)|A_m|} = 0. \quad (A.5)$$

Thus the SQS deviation as Eq. (A.4) also vanishes. The proof is finished.

However, the inverse proposition does not hold. For a configuration whose SQS deviation function is 0, its SAE objective function may not be 0. We found a counterexample structure as a proof. For an equimolar bcc binary AB solid solution, its lattice constant of unit cell is assumed to be 1.0. For a $3\times 3\times 3$ supercell with 54 atoms, if the cutoff radii for $A_2$ and $A_3$ clusters are set to be 1.05, there are two types of $A_3$ clusters. For the larger $A_3$ lattice cluster, the three interatomic distances are 1.0, 1.0 and 1.0 and there are 54 equivalent $A_3$ clusters in the supercell. For the $A_3$ (A, A, A) atom cluster, the expectation of $|A_3(A,A,A)|_\sigma$ is $E(|A_3(A,A,A)|_\sigma) = 0.5^3 \times 54 = 6.75$, which is not an integer. Meanwhile, for any supercell configuration $\sigma$, $|A_3(A,A,A)|_\sigma$ must be an integer. This leads to an intrinsically non-zero objective function for any $\sigma$. However, there are configurations whose SQS deviation function for this $A_3$ cluster is 0, and the SAE objective function is larger than 0. In the SQS method, the value of the cluster function could be negative, so the summation of the deviations may cancel each other, leading to $\Delta\rho(A, \sigma) = 0$.

Comparing with the SQS method, evaluation of the SAE objective function does not require the calculation of the cluster function. As a result, the implementation of the SAE method is much easier than the SQS method. The optimal SAE supercell configuration is also optimal for SQS method, while the inverse proposition does not hold. From the above discussion, we have proved that the SAE method is different from SQS, and the SAE method offers an alternative approach for structural modeling of solid solution other than SQS.